\documentclass[letterpaper]{sig-alternate-pages}

\newfont{\mycrnotice}{ptmr8t at 7pt}
\newfont{\myconfname}{ptmri8t at 7pt}
\let\crnotice\mycrnotice%
\let\confname\myconfname%

\permission{}
\conferenceinfo{MSWiM'13,}{November 3--8, 2013, Barcelona, Spain.}
\copyrightetc{doi:\href{http://dx.doi.org/10.1145/2507924.2507943}{10.1145/2507924.2507943}}
\crdata{978-1-4503-2353-6/13/11\ ...\$15.00.\\
http://dx.doi.org/10.1145/2507924.2507943 }

\makeatletter
\newcommand\footnotescriptsize{\@setfontsize\footnotescriptsize\@vipt{6\p@}}
\makeatother
 
\usepackage{hyperref}
\usepackage{stmaryrd}
\usepackage{enumitem}

\usepackage{breakurl}    
\usepackage{amsfonts,amsmath,amssymb}
\usepackage{algorithm,algorithmic}
\makeatletter
\def\theHALC@line{\thealgorithm-\theALC@line}
\def\theHALC@rem{\thealgorithm-\theALC@rem}
\makeatother
\floatname{algorithm}{Process}

\usepackage{graphicx}
\usepackage{enumerate}
\usepackage{color}
\usepackage{aodv}
\usepackage{xspace}
\renewcommand{\msg}[3]{#1${}_{\!\!\;\MakeUppercase{#2}\!\!\:\shortrightarrow\!\MakeUppercase{#3}}$\xspace}

\newcommand{\gennode}[1]{\ensuremath{\MakeUppercase{#1}}\xspace}
\newcommand{\na}{\gennode{a}}

\newcommand{\nd}{\gennode{d}}
\newcommand{\ns}{\gennode{s}}
\newcommand{\nx}{\gennode{x}}

\newcommand{\rte}{routing table entry\xspace}
\newcommand{\rtes}{routing table entries\xspace}
\newcommand{\uppaal}{\textsc{Uppaal}}

\renewcommand{\sf}{\it}
\begin{document}
\title{Sequence Numbers Do Not Guarantee\\ Loop Freedom\\
---AODV Can Yield Routing Loops---}
\numberofauthors{4}
\author{
\alignauthor Rob van Glabbeek\\
       \affaddr{NICTA, Australia}\\[0.5mm]
       \affaddr{University~of~New~South~Wales,}\\
       \affaddr{Australia}\\
       \affaddr{\tt rvg@cs.stanford.edu}
\alignauthor Peter H{\"o}fner\\
       \affaddr{NICTA, Australia}\\[0.5mm]
       \affaddr{University~of~New~South~Wales,}\\
       \affaddr{Australia}\\
       \affaddr{\tt Peter.Hoefner@nicta.com.au}
\and
\alignauthor Wee Lum Tan\\
       \affaddr{NICTA, Australia}\\[0.5mm]
       \affaddr{University~of~Queensland,}\\
       \affaddr{Australia}\\
       \affaddr{\tt WeeLum.Tan@nicta.com.au}
\alignauthor Marius Portmann\\
       \affaddr{NICTA, Australia}\\[0.5mm]
       \affaddr{University~of~Queensland,}\\
       \affaddr{Australia}\\
       \affaddr{\tt marius@itee.uq.edu.au}
}

\maketitle
\vspace{-24pt} 
\begin{abstract}
In the area of mobile ad-hoc networks and wireless mesh networks,
sequence numbers are often used in routing protocols to avoid routing
loops.  
It is commonly stated in protocol specifications that  sequence numbers are sufficient to
{\em guarantee} loop freedom if they are monotonically increased over
time.  A classical example for the use of sequence numbers is the
popular Ad hoc On-Demand Distance Vector (AODV) routing protocol.  
The loop freedom of AODV is not only 
a common belief, it has been claimed in the abstract of its RFC
and at least two proofs have been proposed.
AODV-based protocols such as AODVv2 (DYMO) and 
HWMP also claim loop freedom due to the same use of sequence numbers.

In this paper we show that AODV is not a priori loop free; 
by this we counter the proposed proofs in the literature.
In fact, loop freedom hinges on non-evident assumptions to be made when
resolving ambiguities occurring in the RFC\@.  Thus, monotonically increasing sequence
numbers, by themselves, do {\em not} guarantee loop freedom.
\end{abstract}
\vfill

\section{Introduction}\label{sec:intro}
Wireless Mesh Networks (WMNs), which can be considered to include Mobile Ad-hoc Networks (MANETs),
have gained considerable popularity and are increasingly deployed in a wide range of application scenarios, including emergency response communication, intelligent transportation systems, mining and video surveillance. They are self-organising wireless multi-hop networks that can provide broadband communication without relying on a wired backhaul infrastructure, a benefit for rapid and low-cost network deployment. 

Highly dynamic topologies are a key feature of WMNs and MANETs,
due to mobility of nodes and/or the variability of wireless links. 
This makes the design and implementation of robust and efficient routing protocols for these networks a challenging task,
and a lot of research effort has gone into~it.

Loop freedom is a critical property for any routing protocol, but it is particularly relevant and challenging for WMNs and MANETs.
Descriptions as in~\cite{Garcia-Luna-Aceves89} capture the common understanding of loop freedom:
``{\sf A routing-table loop is a path specified in the nodes' routing tables at a particular point in time that visits the same node more than once before reaching the intended destination.}"
Packets caught in a routing loop,  until they are discarded by the IP Time-To-Live (TTL) mechanism, can quickly saturate the links and have a detrimental impact on network performance. It is therefore critical to ensure that protocols prevent routing loops.

Sequence numbers, indicating the freshness of routing information, have been widely used to guarantee loop freedom, in particular for distance vector protocols such as DSDV~\cite{DSDV94}, AODV~\cite{rfc3561}, AODVv2 (formerly known as DYMO)~\cite{DYMO25} and HWMP~\cite{HWMP}. 
These  protocols claim to be loop free due to the use of monotonically increasing sequence numbers.
For example, the AODV RFC states:  AODV ``{\sf uses destination sequence numbers to ensure loop freedom at all
 times (even in the face of anomalous delivery of routing control messages), ...}"~\cite{rfc3561}, 
and a similar claim is made in the IETF draft of AODVv2~\cite{DYMO25}:
``{\sf AODVv2 uses sequence numbers to assure loop freedom
  [Perkins99].\/}''\footnote{Here, [Perkins99]
is our reference~\cite{AODV99}.}
A proof of loop freedom of AODV has been provided in~\cite{AODV99}.  
Another, more recent proof is~\cite{ZYZW09}.\footnote{We discuss problems with these proofs later in this paper.}
It is therefore a common belief that the use of sequence numbers in this context guarantees loop freedom. 

However, while this use of sequence numbers can be an efficient approach to address the problem of routing loops, 
we show in this paper that  sequence numbers by themselves do not guarantee loop freedom. 
We illustrate this using AODV as a running example. 

We show that loop freedom can be guaranteed only if sequence numbers are used in a careful way, considering further rules and assumptions on the behaviour of the protocol.
The problem is, as shown in the case of AODV, that these additional rules and assumptions are not explicitly stated in the RFC, 
and that the RFC has significant ambiguities in this regard.
We demonstrate that routing loops can be created---while fully complying with the RFC, and
making reasonable assumptions when the RFC allows different interpretations.
As a consequence, which is a {\em key contribution} of this paper, we obtain
that routing protocols using sequence numbers as described
in~\cite{rfc3561} are not a priori loop free. We argue that the lack
of precision and the corresponding ambiguity of the protocol
definition in the RFC is a key problem here---and for RFCs in general. As {\em another contribution} we show
details of several ambiguities and contradictions in the AODV RFC,
and discuss which interpretations will lead to routing loops.
A {\em third contribution} of this paper is an ana\-lysis of five
key implementations of the AODV protocol, and a discussion of their
corresponding loop freedom properties.

To address the problem of ambiguities and contradictions in
  RFCs,  we argue for the benefit of more precise and formal
approaches for the specification of protocols, which are sufficiently
expressive to model real networks and protocols, while maintaining
usability. As a {\em final contribution} we show how formal methods can be used to avoid ambiguities in RFCs and to guarantee properties such as loop freedom.

The remainder of the paper is organised as follows: 
in Section~\ref{sec:aodv} we recapitulate the basic principles of the AODV protocol.
We state what we mean by loop freedom 
and discuss
the existing proofs of AODV's loop freedom.
In Section~\ref{sec:RFCrevisited}, nearly all plausible
readings of the AODV RFC, resolving its ambiguities and contradictions, are discussed. 
Each of the presented interpretations is analysed w.r.t.\ loop freedom.
In Section~\ref{sec:loops} we show an example of a routing loop occurring in the AODV routing protocol, as a result of following a reasonable and plausible interpretation of the RFC\@.
Having the various RFC interpretations and the loop example in mind, we then analyse five different
key implementations of AODV w.r.t.\ routing loops.
Finally, in \Sect{methodology}, we sketch how 
formal methods---here in the form of process algebra---can
be applied to model routing protocols and verify the presented results on routing loops.
Before we conclude with a short discussion in Section~\ref{sec:conclusion},
we present major related work in Section~\ref{sec:related}.

\section{AODV}\label{sec:aodv}
To prove that sequence numbers do not a priori guarantee loop freedom, 
we use AODV as an example. We expect that other routing protocols such
as AODVv2 and HWMP behave similarly.
AODV~\cite{rfc3561} is a popular routing protocol designed for
MANETs, and is one of the four protocols currently standardised by the
IETF MANET working group\footnote{\url{http://datatracker.ietf.org/wg/manet/charter/}}.
It also forms the basis of new WMN routing protocols, including the
upcoming IEEE 802.11s wireless mesh network standard~\cite{HWMP}.
AODV is designed for wireless and mobile networks where links are particularly 
unreliable.

\subsection{Brief Overview}\label{ssec:aodv}

AODV is a reactive protocol, which means that routes are established only
on demand.  If a node~\ns wants to send a data packet to  a node \nd, but
currently does not know a route, it temporarily buffers the
packet and initiates a route discovery process
by broadcasting a route request (RREQ) message in the network. An intermediate
node \na that receives the RREQ message creates a \rte for a route
towards node~\ns referred to as a \emph{reverse route}, and re-broadcasts the RREQ. This is repeated until the RREQ reaches the destination node~\nd, or alternatively a node that knows a route to \nd. In both cases, the node replies by unicasting a corresponding route reply (RREP) message back to the source~\ns,
via a previously established reverse route. When forwarding RREP messages, nodes
create a \rte for node \nd, called the \emph{forward route}. When the RREP reaches
the originating node \ns, a route from \ns to \nd  is established and data packets
can start to flow. Both forward and reverse routes are maintained in a routing table at every node---details are given below.
In the event of link and route breaks, AODV uses route error (RERR) messages to
notify the affected nodes: if a link break is detected by a node, it first invalidates all routes stored in the node's own routing table that actually use the broken link. Then it sends a RERR message containing the unreachable destinations to all (direct) neighbours using this route. 

In AODV, a routing table consists of a list of entries---at most one for each destination---each containing the following information:

\begin{itemize}[leftmargin=13.2pt]
\setlength{\itemsep}{-1pt}
\item the destination IP address;
\item the destination sequence number;
\item the sequence-number-status flag---tagging whether the recorded sequence number
can be trusted;
\item a flag tagging the route as being valid or invalid---this flag is set to invalid
when a link break is detected;
\item the hop count, a metric to indicate the distance to the destination;
\item the next hop, an IP address that
identifies the next (intermediate) node on the route to the destination;
\item a list of precursors, a set of IP addresses of those
  $1$-hop neighbours that use this particular route; and
\item the lifetime (expiration or deletion time) of the route.
\end{itemize}
The destination sequence number constitutes a measure approximating the
relative freshness of the information held---a higher number denotes newer information.  
The  routing table  is updated whenever a node receives an AODV control message (RREQ, RREP or RERR)
or detects a link break. 

During the life time of the network, each node not only maintains its routing table, it also stores
its {\em own} {\em sequence number}.  This number is used as a local ``timer'' and is incremented 
whenever a new route request is initiated. 

Full details of the protocol are outlined in the request for comments (RFC)~\cite{rfc3561},
the official specification of AODV.

\subsection{Loop Freedom}\label{ssec:loopfreedom}
The ``naive'' notion of loop freedom is a term that informally means that ``a packet never goes round in cycles without (at some point) being delivered''. This dynamic definition is too restrictive a requirement for AODV. There are situations where packets are sent in cycles, but which should not be considered ``harmful''. This can happen when the network topology keeps changing. 
The sense of loop freedom is much better captured by a 
static invariant, saying that {\em at any given time the collective routing tables of the nodes 
do not admit a loop}. Such a requirement does not rule out the dynamic loop 
alluded to above. However, in situations where the topology remains stable sufficiently long it does 
guarantee that packets will not keep going around in cycles.

\subsection{Proofs Based On Sequence Numbers}\label{ssec:sqns}

As mentioned
before, AODV ``{\sf  uses destination sequence numbers to ensure loop freedom at all times\/}''~\cite[Page~2]{rfc3561}.
Moreover, it has been ``proven'' at least twice that AODV is loop free~\cite{AODV99,ZYZW09}. 

In both papers a main argument is that messages relating to a particular route request are handled only once at every node; 
moreover that every route discovery process has a unique sequence number. 
The latter is guaranteed since a node's own sequence number is incremented whenever a new route discovery is initiated.
Each sequence number stored in any routing table for destination $\dval{dip}$ is ultimately derived from
\dval{dip}'s own sequence number at the time such a route was discovered.

The proof sketch given in~\cite{AODV99} uses the fact that when a loop in a route to a destination
$Z$ is created, all nodes $X_i$ on that loop must have route entries for destination $Z$
with the same destination sequence number. ``{\sl Furthermore, because the destination sequence
numbers are all the same, the next hop information must have been derived at every node
$X_i$ from the same RREP transmitted by the destination $Z$\/}''~\cite[Page 11]{AODV99}. The latter is not true at all:
some of the information could have been derived from RREQ messages, or from a RREP message transmitted
by an intermediate node that has a route to $Z$. More importantly, the nodes on the loop
may have acquired their information on a route to $Z$ from different RREP or RREQ
messages, that all carried the same sequence number. This will be illustrated in our
forthcoming loop example.

The proof of~\cite{ZYZW09} was established in two steps: 
(i) a mathematical model of AODV was derived\footnote{In \cite{ZYZW09} the model is only given partially.};
(ii) based on the derived model loop freedom was proven using the interactive theorem prover Isabelle~\cite{NipkowPaulsonWenzel02}.\footnote{Again only snippets of the proofs can be found in~\cite{ZYZW09}.} 
In the model of \cite{ZYZW09} route replies generated by
  intermediate nodes \cite[Sect.\ 6.6.2.]{rfc3561} are not
  considered. Since this is a key feature of AODV (and, as we shall
  see, an essential ingredient in the creation of routing loops), the
  work of \cite{ZYZW09} cannot be said to pertain to the full protocol.
  
\newcommand{\intref}[2]{\ref{int#1}\ref{int#1#2}}
\begin{table*}\
\vspace*{-2ex}
\caption{Analysis of different interpretations of the RFC 3561 (AODV)}
\vspace*{1.5ex}
\centering
\newcounter{tabcounti}
\newcounter{tabcountii}
\setcounter{tabcounti}{0}
\makeatletter
\renewcommand{\p@tabcounti}{}
\renewcommand{\p@tabcountii}{\thetabcounti}
\makeatother
\renewcommand{\thetabcountii}{\alph{tabcountii}}
\newcommand{\nri}[1]{\setcounter{tabcountii}{0}\refstepcounter{tabcounti}\label{int#1}\thetabcounti.}
\newcommand{\nrii}[1]{\refstepcounter{tabcountii}\label{int#1}\thetabcounti\thetabcountii.}
{\small
\setlength\doublerulesep{4em}
\begin{tabular}{@{\,}l@{\,}|p{0.43\textwidth}|p{0.43\textwidth}@{}|@{}}
\hline
\multicolumn{3}{|@{\,}l|}{\bf\nri{1} Updating the Unknown (Invalid) Sequence Number in Response to a Route Reply}\\
\hline
\nrii{1a}			&the destination sequence number (DSN) is copied from the RREP message (Sect 6.7)&may decrement  sequence numbers, which causes loops\\
\cline{2-3}
\nrii{1b}			&the routing table is not updated when the information that it has is ``fresher'' (Sect.\ 6.1) & does not cause loops\\
\cline{2-3}
\multicolumn{3}{c}{}\\[-2.1ex]
\hline
\multicolumn{3}{|@{\,}l|}{
    \bf \nri{2} Updating with the Unknown Sequence Number (Sect.\ 6.5)}\\
\hline
\nrii{2a}			&no update occurs & does not cause loops, but opportunity to improve routes\newline is missed. 
\\
\cline{2-3}
\nrii{2b}			&overwrite any \rte by an update with an unknown DSN& may decrement  sequence numbers, which causes loops\\
\cline{2-3}
\nrii{2c}			&use the new entry with the old DSN & does not cause loops\\
\cline{2-3}
\multicolumn{3}{c}{}\\[-2.1ex]%
\hline
\multicolumn{3}{|@{\,}l|}{
    \bf \nri{3} (Dis)Allowing the Creation of Self-Entries in Response to a Route Reply}\\
\hline
\nrii{3a}				&allow (arbitrary) self-entries& loop free only if used with Interpretations 4d or 4e below\\
\cline{2-3}
\nrii{3b}				&allow optimal self-entries only;\newline store own sequence number in optimal self-entry &does not cause loops\\
\cline{2-3}
\nrii{3c}				&disallow self-entries; if self-entries would occur, ignore msg.&does not cause loops\\
\cline{2-3}
\nrii{3d}				&disallow self-entries; if self-entries would occur, forward &does not cause loops\\
\cline{2-3}
\multicolumn{3}{c}{}\\[-2.1ex]
\hline
\multicolumn{3}{|@{\,}l|}{
    \bf \nri{4} Invalidating Routing Table Entries in Response to a Route Error Message}\\
\hline    
\nrii{4a}				&always copy DSN from RERR message (Sect.\ 6.11)&may decrement  sequence numbers, which causes loops\newline
						(when allowing self-entries (Interpretation 3a))\\
\cline{2-3}
\nrii{4b}				&only invalidate if the DSN in the routing table is smaller than or equal to the one from the RERR message (Sect.~6.1)
 						&causes loops (when allowing self-entries) \\
\cline{2-3}
\nrii{4c}				&take the maximum of the DSN of the routing table and the one from the RERR message 
						&causes loops (when allowing self-entries)
		\\
\cline{2-3}
\nrii{4d}				&take the maximum of the increased DSN of the routing table and the one from the RERR message
 						&does not cause loops
 \\
\cline{2-3}
\nrii{4e}				&only invalidate if the DSN in the routing table is smaller than the one from the RERR message (Sect.~6.2) 
						&does not cause loops\\
\cline{2-3}
\end{tabular}
}
\label{tab:interpret}
\vspace*{-1ex}
\end{table*}

\section[The AODV RFC]{Ambiguities in the AODV RFC}\label{sec:RFCrevisited}

``{\sf  
A Request for Comments (RFC) is a publication of the Internet Engineering Task Force (IETF) and the Internet Society {[\dots]}.
A RFC is authored by engineers and computer scientists in the form of a memorandum describing methods, behaviors, research, or innovations applicable to the working of the Internet and Internet-connected systems.
{[\dots]} The IETF adopts some of the proposals published as RFCs as Internet standards.\/}''\footnote{\url{http://en.wikipedia.org/wiki/Request_for_Comments}}
Not all RFCs are standards~\cite{rfc1768}. However, in the case of the AODV routing protocol, the RFC 3561~\cite{rfc3561} is the de facto standard.

\hspace{-1.28546pt}%
RFCs are typically written in English and are not equipped with a formal description language. 
This has the advantage that everybody can read any RFC\@. 
However, it holds the disadvantage that the RFC contains contradictions and ambiguities. 
Additionally, there may be unexpected situations occurring in real
networks that are not described or anticipated by the RFC\@. 
In sum, this yields a variety of interpretations for every RFC\@; some interpretations being more plausible than others.

In this section, we discuss some of the ambiguities found in the specification of AODV.
We catalogue and analyse the variants of AODV
arising from interpretations consistent with the reading of the RFC and show which of them yields routing loops.\
Our analysis is based on a rigorous, formal, and mathematical approach~\cite{ESOP12}.
A full and detailed analysis is given in~\cite{TR11}.
\Tab{interpret} summarises the results; section numbers refer to~\cite{rfc3561}.

One of the crucial aspects of AODV is the maintenance of routing tables.
Hence the update of \rtes with new information has to be performed carefully. 
Unfortunately, the RFC specification only gives hints 
how to update \rtes; an exact and precise definition is missing.

\newcounter{myp}
\setcounter{myp}{1}
\newcommand{\myparagraph}[1]{{\vspace{0pt plus 4pt}\par\hspace{-3pt}\textbf{\arabic{myp}.~#1.}\addtocounter{myp}{1}}}

\myparagraph{Updating the Unknown (Invalid) Sequence Number in Response to a Route Reply}
If a node {receives a RREP message, it might have to update its routing table}:
``{\sf
  the existing entry is updated only in the following circumstances:
  (i) the sequence number in the routing table is marked as invalid {[\dots]}\/}''~\cite[Sect.~6.7]{rfc3561}.
Here it is relevant that, through the sequence-number-status flag, any
  sequence number can be marked as unknown or invalid.
 In the same section it is also stated what actions occur if a route is updated. Among others it is stated that 
 ``{\sf the destination sequence number is marked as valid, [\dots] and the [new] destination sequence number [in the routing table] is the Destination Sequence 
      Number in the RREP message.\/}''~\cite[Sect.\ 6.7]{rfc3561}.
 The interpretation that follows these lines literally is denoted \ref{int1a} in Table~\ref{tb:analysis}. It
can decrement sequence numbers,
which immediately yields routing loops, as explained in \cite[Sect.~8]{TR11}.
This update mechanism contradicts Sect.\ 6.1 of \cite{rfc3561},
  which states that any information from an incoming AODV control
  message that carries a lower sequence number than the corresponding
  entry in the routing table MUST be discarded (Interpretation \ref{int1b}).
  The routing loops resulting from following Sect.\ 6.7 strongly
  indicate that this contradiction in the RFC should be resolved in
  favour of Sect.\ 6.1.

\myparagraph{Updating with the Unknown Sequence Number}
Whenever a node receives a forwarded AODV control message from a $1$-hop neighbour, it creates a new or updates an existing \rte to that neighbour.
For example, ``{\sf [w]hen a node receives a RREQ, it first creates or updates a route to
   the previous hop without a valid sequence number\/}''~\cite[Sect.~6.5]{rfc3561}.
In case a new \rte is created, the sequence number is set to a default value (typically $0$, as is done in implementations such as AODV-UU \cite{AODVUU}, AODV-UIUC \cite{Kawadia03} and AODV-UCSB \cite{CB04}) and the sequence-number-status flag is set to {\unkno} to signify that the sequence number corresponding to the neighbour is unknown. 
But, what happens if there exists already a \rte?
Following the quote above, the routing table has to be updated. Unfortunately, it is not stated 
how the update is done.\footnote{Section 6.2 of \cite{rfc3561} further explains in which
  circumstances an update occurs. It does not resolve this ambiguity (cf.~\cite{TR11}).}
There are three reasonable options:\\
(\ref{int2a}) no update occurs.
This interpretation is harmless with regard to routing loops, but misses an opportunity to improve some routes.
It can be argued that the RFC rules out this option by including ``or updates'' in the quote above.\\
(\ref{int2b}) All information is taken from the incoming AODV control message;
since that message formally does not contain a sequence number for the neighbour, the destination
sequence number is set to value $0$.
Since this can decrease sequence numbers, 
routing loops might occur. Hence this interpretation must not be used.\\
(\ref{int2c}) The information from the routing table
and from the incoming AODV control message is
merged.\footnote{By taking the destination sequence number from the
  existing \rte and all other information from the AODV
  control message.}
This interpretation does not give rise to 
loops.

\myparagraph{(Dis)Allowing the Creation of Self-Entries in Response to a Route Reply}
In any practical implementation, when a node sends a data packet to itself, 
the packet will be delivered to the corresponding application on the 
local node without ever involving a routing protocol and therefore without being ``seen'' 
by AODV or any other routing protocol. 
Because of this it seems that it does not make a difference whether any node using AODV  stores \rtes to itself.

In AODV, when a node receives a RREP message, it creates a routing table entry for the
  destination node if such an\pagebreak[4] entry does not already exist \cite[Sect. 6.7]{rfc3561}. If the destination node
  happens to be the processing node itself, this leads to the creation of a self-entry. The RFC does
  not mention self-entries explicitly; it only refers to them at one location:
  ``{\sf 
  A node may change the sequence number in the routing table entry of a 
   destination only if: --  it is itself the destination node {[\dots]}\,}''~\cite[Sect. 6.1]{rfc3561}.
This points at least to the possibility of having self-entries. We have analysed various implementations of AODV and found that the Kernel-AODV \cite{AODVNIST}, AODV-UIUC \cite{Kawadia03}, AODV-UCSB \cite{CB04} and AODV-ns2 implementations allow the creation of self-entries.

If arbitrary self-entries are allowed (Interpretation \ref{int3a} in Table~\ref{tb:analysis}) this can, in combination with other plausible
assumptions, yield routing loops, as we will show in the next section. 
However, storing only optimal self-entries  in routing tables (Interpretation \ref{int3b}) does not cause loops.
For example, Kernel-AODV maintains the nodes' own sequence numbers in this way.

On the other hand, there are two possibilities to disallow self-entries: 
if a node receives a route reply and would create a self-entry, it silently discards the message (Interpretation \ref{int3c}). 
This interpretation has the disadvantage that replies are lost. 
The alternative is that  the node forwards the message without updating its routing table (\ref{int3d}).
Both variants by themselves do not yield routing loops. 

\myparagraph{Invalidating Routing Table Entries in Response to a Route Error Message}
If a node receives a RERR message, it might invalidate entries of its routing table based on information 
from this message. When invalidating routing table entries, destination sequence numbers
should be ``{\sf copied from the incoming RERR\/}''~\cite[Sect.~6.11]{rfc3561}.
This is Interpretation \ref{int4a} in Table~\ref{tab:interpret}.
In particular, this part of the RFC prescribes the replacement of an
existing destination sequence number in a \rte with one that
may be strictly smaller, which contradicts Sect.~6.1 of the RFC\@.

To make 
invalidation consistent with Sect.~6.1 of the RFC, one could
use two possible variants instead. The first (\ref{int4b}), strictly following Sect.~6.1, 
invalidates only if the destination sequence number in the routing table is smaller than or equal to 
 the destination sequence number provided by the incoming RERR message; it aborts otherwise.
The second (\ref{int4c})
invalidates in all circumstances, but prevents a decrease in the destination
sequence number by taking the maximum of the stored and the incoming number.%
\footnote{Although this is not really a plausible reading of the RFC, Interpretation  \ref{int4c} can be seen as a compromise between Sections 6.11 and 6.1 of the RFC---it seems a natural way to avoid a decrement of destination sequence numbers and still 
take all information from the RERR message into account.}
In the next section we will show that each of these three interpretations can yield 
routing loops, when used in conjunction with non-optimal self-entries.

There are two reasonable solutions to avoid routing loops in these circumstances. 
As a modification of Interpretation \ref{int4c},
one can first increment the destination sequence number of the routing
table by one and then use the maximum of this updated sequence number
and the one from the RERR message (Interpretation \ref{int4d}).
Alternatively, one could 
invalidate only if the destination sequence number in the routing table is (strictly) smaller than the destination sequence number provided by the incoming RERR 
message (\ref{int4e}),
and abort if it is larger or equal.
The latter interpretation is at least consistent with Section 6.2 of the RFC, which states that
``{\em The route is only updated if  the new sequence number is either (i)
   higher than the destination sequence number in the route table, or {[\dots]\/}}''.

\section{AODV Yields Loops}\label{sec:loops}
In the previous section, we discussed some of the ambiguities found in the specification of AODV, and catalogued the variants of AODV that arise from interpretations consistent with the RFC.
In the following we describe an example of a routing loop in AODV, based on a reasonable and
plausible interpretation of the RFC, following Interpretations \ref{int3a} and any out of \ref{int4a}, \ref{int4b} and \ref{int4c}
  of AODV, in resolving the ambiguities from \Tab{interpret}.

\subsection{Creating Routing Loops}\label{ssec:self_to_loop}

\begin{figure*}
{\small
\newcommand{\myscale}{1.15}  
\newcommand{\hi}{\hspace{15pt}}
\newcommand{\hii}{\hspace{27pt}}
\centering
\begin{tabular}{|@{~}p{86mm}@{~\,}|@{~}p{86mm}@{~}|}\hline
\rule{0mm}{3mm}
(a) The initial state.&(b) Standard RREQ-RREP cycle; \nd is looking for \na;\\
&\textcolor{white}{(b) }\na, \nd establish routing table entries for each other.\footnotemark\\[1ex]
\hii\includegraphics[scale=\myscale]{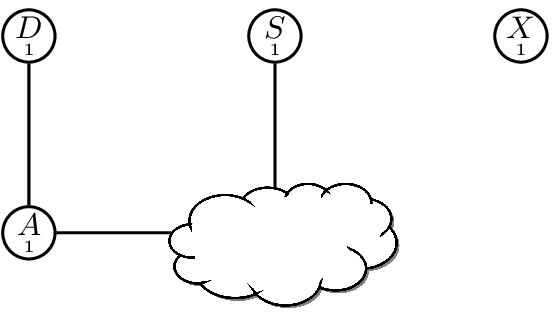}&
\hi\includegraphics[scale=\myscale]{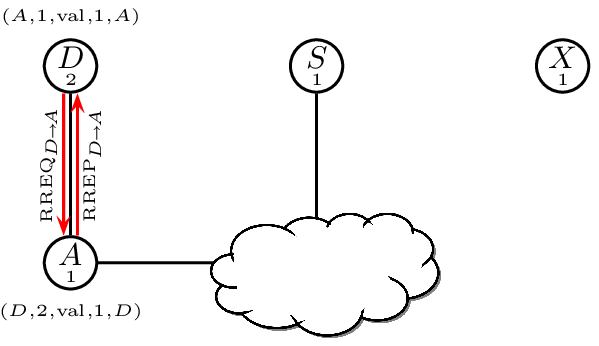}\\
\hline
\rule{0mm}{3mm}
(c) \ns broadcasts RREQ message searching for \nd.&(d) Topology changes;\\
&\textcolor{white}{(d) }\ns searches for \nx; RREQ message flows throughout the network.\\[1ex]
\hi\includegraphics[scale=\myscale]{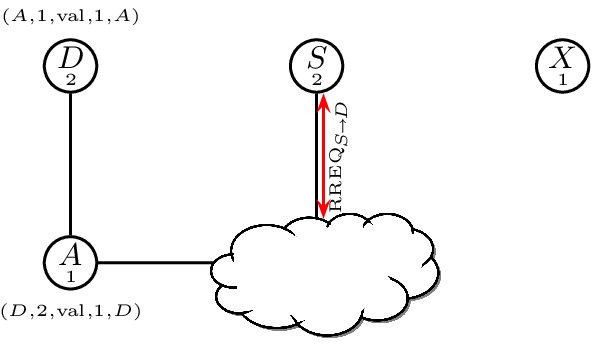}&
\hi\includegraphics[scale=\myscale]{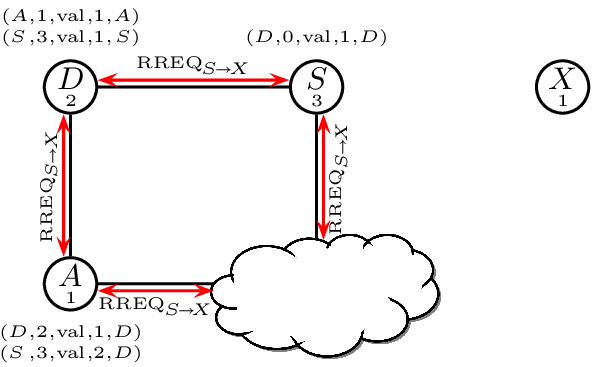}\\
\hline
\rule{0mm}{3mm}
(e) Finally \msg{RREQ}{s}{d} (initiated in Part (c)) reaches node \na;&(f) Topology changes; Standard RREQ-RREP cycle;\\
\textcolor{white}{(e) }\na processes it and unicasts a reply back (via \nd);&\textcolor{white}{(f) }\nd initiates route request searching for \nx.\\
\textcolor{white}{(e) }Node \nd establishes self-entry.&\\
\hi\includegraphics[scale=\myscale]{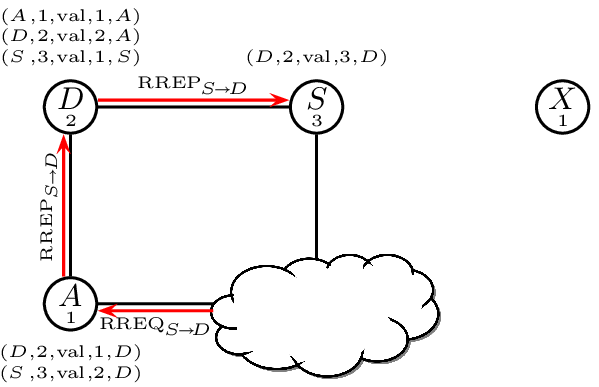}&
\hi\includegraphics[scale=\myscale]{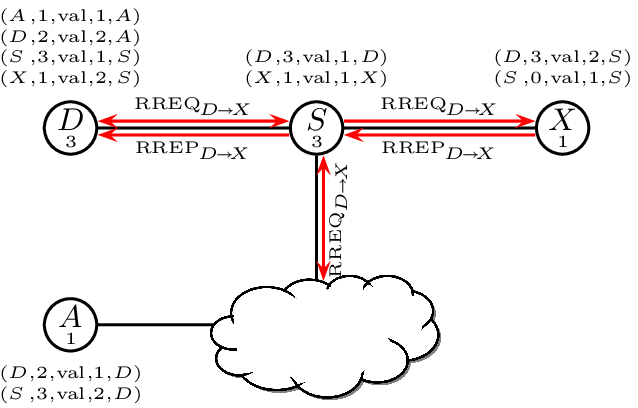}\\
\hline
(g) Topology changes;&(h) Topology changes;\\
\textcolor{white}{(g) }Error message is generated by \nd.&\textcolor{white}{(h) }\ns initiates route request searching for \nd.\\[1ex]
\hi\includegraphics[scale=\myscale]{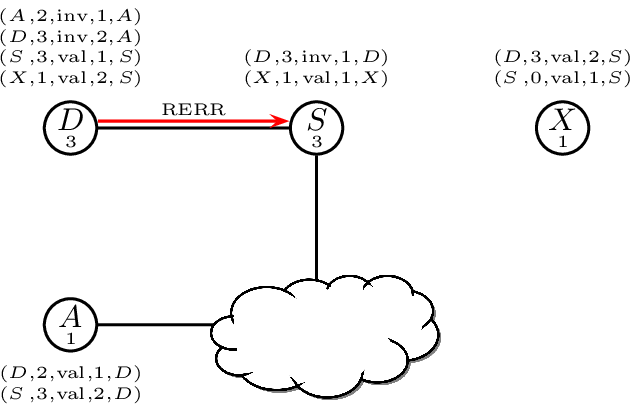}&
\hi\includegraphics[scale=\myscale]{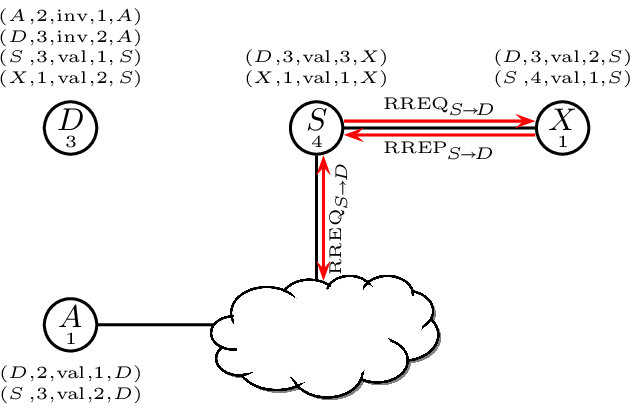}\\
\hline
    \end{tabular}
}
\caption{Creating routing loops}
\vspace*{5mm}
\footnoterule
\addtocounter{footnote}{-1}
{\footnotesize  \footnotemark
Each routing table entry has the form $(\textit{destination},\textit{sequence number},\textit{validity},\textit{hop count},\textit{next hop})$.
}
\label{fig:routingloop}
\end{figure*}

The given example (shown in Figure~\ref{fig:routingloop}) consists of four parts:
(1) First, a standard RREQ-RREP cycle occurs (Figures~\ref{fig:routingloop}(a)--(b));
(2) Then, a node stores information about {\em itself} in its routing table (Figures~\ref{fig:routingloop}(c)--(e)); 
Such information is called a {\em self-entry}.
(3) Another standard RREQ-RREP cycle occurs (Figure~\ref{fig:routingloop}(f));
(4) Finally, a link break in combination with another route discovery yields the
loop (Figures~\ref{fig:routingloop}(g)--(h)).

\setcounter{myp}{1}

\myparagraph{\textbf{Standard RREQ-RREP Cycle}}
In Figure~\ref{fig:routingloop}(a), we show the initial network topology, with the nodes' sequence numbers depicted inside the circles. Figure~\ref{fig:routingloop}(b) shows node \nd searching for a route to node \na. We see that the sequence number for node \nd is increased to $2$ (``{\sf The Originator Sequence Number in the RREQ message is the
node's own sequence number, which is incremented prior to insertion in a RREQ.\/''}~\cite[Sect. 6.3]{rfc3561}).
Figure~\ref{fig:routingloop}(b) also shows the nodes' routing tables. Each \rte (depicted as a
$5$-tuple) contains information about the destination, the destination's sequence number, the
validity of the \rte, the hop count and the next hop (cf.\ Section~\ref{ssec:aodv}). We do not show
the sequence-number-status flag, the list of precursors in the \rtes and the lifetime of a route as they constitute auxiliary information that is not critical to the loop example here. Due to the successful exchange of RREQ-RREP messages, nodes \nd and \na create \rtes to each other in their routing table.

\myparagraph{\textbf{Self-Entries}}
In Figure~\ref{fig:routingloop}(c), we see node \ns searching for a route to node \nd. In Figure~\ref{fig:routingloop}(d), the link between nodes \ns and \nd goes up (e.g.\ due to node mobility) and node \ns then searches for a route to node \nx. The route request message \msg{RREQ}{s}{x} is forwarded by nodes \nd and \na since they do not have any information on the destination node \nx. From the information contained in the \msg{RREQ}{s}{x} message, a \rte to node \ns is created in the routing tables of nodes \nd and \na. 
``{\sf Then the node searches for a reverse route to the Originator IP Address {[\dots]}  If need be, the route is created, or updated using the Originator Sequence Number from the RREQ in its routing table.\/}''~\cite[Sect. 6.5]{rfc3561}. 
Node \ns also receives the forwarded \msg{RREQ}{s}{x} message from node \nd, and before silently discarding the message (since it is the originator of the RREQ message), updates its routing table to create an entry to node \nd.
``{\sf When a node receives a RREQ, it first creates or updates a route to the previous hop without a valid sequence number.\/}''~\cite[Sect. 6.5]{rfc3561}. We use the value $0$ for an unknown/invalid sequence number created in this manner, as is also done in implementations like AODV-UU~\cite{AODVUU}, AODV-UIUC \cite{Kawadia03} and AODV-UCSB \cite{CB04}.

At some point, \msg{RREQ}{s}{d} finally reaches node \na (Figure~\ref{fig:routingloop}(e)). Since node \na has a valid \rte to node \nd, it generates an intermediate route reply message using the information from its routing table ~\cite[Sect. 6.6 and 6.6.2]{rfc3561}. 
The \msg{RREP}{s}{d} message is unicast to node \nd, the next hop on the path towards node \ns. ``{\sf Once created, the RREP is unicast to the next hop toward the originator of the RREQ, as indicated by the \rte for that originator.\/}''~\cite[Sect. 6.6]{rfc3561}. Node \nd processes the \msg{RREP}{s}{d} message, updates its routing table and 
forwards the message to node \ns, which establishes a route to node \nd. When updating its routing table, node \nd creates a self-entry (following Interpretation 3a in \Tab{interpret}) since the \msg{RREP}{s}{d} message contains information about a route to node \nd ~\cite[Sect. 6.7]{rfc3561}. 

\myparagraph{\textbf{Standard RREQ-RREP Cycle}}
In Figure~\ref{fig:routingloop}(f), the link between nodes \ns and \nx goes up, while the link between nodes \nd and \na goes down. Node \nd also searches for a route to node \nx. Due to the successful exchange of RREQ-RREP messages, the routing tables of nodes \nd, \ns, and \nx are updated accordingly.

\myparagraph{\textbf{From Self-Entries to Loops}}
The loop example continues with the link between nodes \ns and \nx going down (Figure~\ref{fig:routingloop}(g)). In addition, node \nd detects that its link to node \na is broken. Following from this, node \nd initiates processing for a RERR message. 
``{\sf A node initiates processing for a RERR message {[\dots]} if it detects a link break for the next hop of an active [(\val)] 
route in its routing table while transmitting data\/}''~\cite[Sect. 6.11]{rfc3561}.
In this process, ``{\sf the node first makes a list of unreachable destinations consisting of the unreachable neighbor and any additional destinations {[\dots]} in the local routing table that use the unreachable neighbor as the next hop.}"~\cite[Sect. 6.11]{rfc3561}. In addition, the routing table for node \nd has to be updated for these unreachable destinations as follows:
``{\sf 
1.~The destination sequence number of this route entry, if it exists and is valid, is incremented {[\dots]}. 
2.~The entry is invalidated by marking the route entry as invalid\/}''.~\cite[Sect. 6.11]{rfc3561}. 
The result of this process is that the sequence numbers of \rtes for the unreachable destination nodes \na and \nd are increased, and the entries invalidated, as shown in the routing table of node \nd in Figure~\ref{fig:routingloop}(g).

The RERR message generated by node \nd contains information about the unreachable destination nodes \na and \nd, taken from the routing table of node \nd. The message is sent to node \ns since it is the precursor for the unreachable destination node \nd.
Node \ns receives the RERR message, and updates its routing table as follows:
``{\sf 1.~The destination sequence number of this routing entry {[\dots]} is copied from the incoming RERR. {[\dots]} 2.~The entry is invalidated by marking the route entry as invalid\/}''.~\cite[Sect. 6.11]{rfc3561}.
Therefore, the entry to node \nd in node \ns's routing table is updated to $(D,3,\inval,1,D)$.%
\footnote{Each of the Interpretations \ref{int4a}, \ref{int4b} and \ref{int4c} in \Tab{interpret} will produce the same update in this scenario.}

In Figure~\ref{fig:routingloop}(h), the link between nodes \ns and \nd goes down, while the link between nodes \ns and \nx goes up. Node \ns also searches for a route to node \nd. Accordingly, the destination sequence number in the \msg{RREQ}{s}{d} message is set to the value $3$ since ``{\sf a previously valid route to the destination {[\dots]} is marked as invalid. {[\dots]} The Destination Sequence Number field in the RREQ message is the last known destination sequence number for this destination and is copied from the Destination Sequence Number field in the routing table.\/}''~\cite[Sect. 6.3]{rfc3561}. The RREQ message is received by node \nx, which generates an intermediate RREP message since ``{\sf it has an active route to the destination, the destination sequence number in the node's existing route table entry for the destination is valid and greater than or equal to the Destination Sequence Number of the RREQ\/}''~\cite[Sect. 6.6]{rfc3561}. 
Due to this, node \nx unicasts a \msg{RREP}{s}{d} message back to node \ns. Finally, node \ns receives this message and updates its routing table, as shown in Figure~\ref{fig:routingloop}(h).

A routing loop between nodes \ns and \nx for destination node \nd is now established.
When either of the nodes has a data packet to send to destination node \nd, the data packet will loop between the two nodes.

In this section, we have shown that AODV is not a priori loop free. 
The presented example can create loops if
(i) sequence numbers are implemented in a way consistent with the RFC,
(ii) self-entries are allowed, and 
(iii) destination sequence numbers are copied directly from RERR messages%
---even when this copying is only executed if it does not cause a decrement in the destination sequence number in the routing table.
This shows that loop freedom hinges on non-evident assumptions to be made when interpreting the RFC 
and not only on monotonically increasing sequence numbers.

\section{AODV Implementations}\label{sec:implementation}
To show that our results are not only theoretically driven, but {\em do} occur in practice, 
we analyse five different open source implementations of AODV:
\begin{itemize}[leftmargin=8.6pt]
\setlength{\itemsep}{-1pt}
\item[$\bullet\!$] {\em AODV-UU}~\cite{AODVUU} is an implementation of AODV, developed at Uppsala University.\\
{\url{http://aodvuu.sourceforge.net/}}
\item[$\bullet\!$] {\em Kernel-AODV}~\cite{AODVNIST} is an implementation developed at NIST\!.\\
 {\url{http://w3.antd.nist.gov/wctg/aodv_kernel/}}
\item[$\bullet\!$] {\em AODV-UIUC}~\cite{Kawadia03} (Univ.~of Illinois at Urbana-Champaign) is an implementation that is based on an early draft (version 10) of AODV.\\ {\url{http://sourceforge.net/projects/aslib/}}
\item[$\bullet\!$] {\em AODV-UCSB}~\cite{CB04} (Univ.~of California, Santa Barbara) is another implementation that is based on an early draft (version 6) of AODV.\hspace{6mm}
\url{http://moment.cs.ucsb.edu/AODV/aodv-ucsb-0.1b.tar.gz}
\item[$\bullet\!$] {\em AODV-ns2\/} is an AODV implementation in the ns2 network simulator, originally developed by the CMU Monarch project and improved upon later by S.~Das and E.~Belding-Royer (the authors of the AODV RFC \cite{rfc3561}). It is based on an early draft (version 8) of AODV. It is frequently used by academic and industry researchers to simulate AODV. \\
\parbox{7.4cm}{\url{http://ns2.sourcearchive.com/documentation/2.35~RC4-1/aodv_8cc-source.html}}
\end{itemize}
Although these implementation behave differently, all of\linebreak them {\em do} capture
the main aspects of the AODV protocol, as specified in the RFC~\cite{rfc3561}. As we have shown in
the previous sections, implementing the AODV protocol based on the RFC specification does not
necessarily guarantee loop freedom: routing loops may occur when following either
  Interpretation \ref{int1a}, Interpretation \ref{int2b} or the combination of Interpretation \ref{int3a} with any of \ref{int4a}--c of \Tab{interpret}.
Therefore, we look at these five concrete AODV implementations to determine whether any of them is susceptible to routing loops. In particular, we examine the code of these implementations to see if routing loops
can occur. 
Table~\ref{tb:analysis} shows the results of this analysis: it
indicates for each of the implementations which of the interpretations of Table~\ref{tab:interpret} they follow and whether they can create routing loops or not.

We found that none of the five implementations makes use of the
sequence-number-status flag, so a positive sequence
number can never be marked as unknown.
As a result, Ambiguity 1 of \Tab{interpret} does not arise.

\begin{table*}
\caption{Analysis of AODV implementations}
\vspace*{1.5ex}
\centering
{\footnotesize
\setlength{\tabcolsep}{2.6pt}
\begin{tabular}{@{}|l|r|l|@{}}
\hline
{\bf Implementation} & {\bf Interpretation} & {\bf Analysis}\\
\hline\hline
\rule[9pt]{0pt}{0pt}%
AODV-UU~\cite{AODVUU} 		& \hfill \ref{int2c},\hfill \ref{int3c},\hfill \ref{int4a} &Loop free, since self-entries are explicitly excluded.\\[1pt]
\hline
\rule[9pt]{0pt}{0pt}%
Kernel-AODV~\cite{AODVNIST} 	&\hfill \ref{int2a},\hfill \ref{int3b},\hfill \ref{int4a} &Loop free, due to optimal self-entries.\\[1pt]
\hline
\rule[9pt]{0pt}{0pt}%
AODV-UIUC~\cite{Kawadia03} 	& \hfill \ref{int2b},\hfill \ref{int3a},\hfill \ref{int4a} &Yields loops, through decrement of sequence numbers, by use of Interpretation \ref{int2b}.\\[1pt]
\hline
\rule[9pt]{0pt}{0pt}%
AODV-UCSB~\cite{CB04}		& \hfill \ref{int2b},\hfill \ref{int3a},\hfill \ref{int4b} &Yields loops, through decrement of sequence numbers, by use of Interpretation \ref{int2b}.\\[1pt]
\hline
\rule[14pt]{0pt}{0pt}%
AODV-ns2		&\hfill \ref{int2a},\hfill \ref{int3a},\hfill \ref{int4b} &\begin{tabular}{@{}l@{}}Yields routing loops in the way described in \Sect{loops},\\
 following plausible interpretations of the RFC
 w.r.t.\ Ambiguities \ref{int3} and \ref{int4}.
 \end{tabular}\\[6pt]
\hline
\end{tabular}}
\label{tb:analysis}
\vspace{-0.5mm}
\end{table*}
In AODV-UU, self-entries are never created because a check is always performed on an incoming RREP
message to make sure that the destination IP address is not the same as the node's own IP
address. In terms of \Tab{interpret} it follows Interpretations \ref{int2c}, \ref{int3c} and \ref{int4a}.
It is shown in~\cite{TR11} that this interpretation of the RFC (avoiding self-entries) is loop free. 

In Kernel-AODV, which follows Interpretations \ref{int2a}, \ref{int3b} and \ref{int4a} of \Tab{interpret}, an optimal self-entry (with hop count 0 and next hop being the node itself) is always maintained by every node in the network. The optimal self-entry is created during node initialisation. 
The node also maintains its own sequence number in this entry.
Since the self-entry is already optimal, a node will never update the self-entry when processing any incoming RREP messages that contain information about itself. As such, a routing loop as described by the example in \SSect{self_to_loop} will never occur.

AODV-UIUC follows the Interpretations~\ref{int2b}, \ref{int3a} and \ref{int4a},
whereas AODV-UCSB implements~\ref{int2b}, \ref{int3a} and \ref{int4b}.
Due to Interpretation \ref{int3a}, both implementations allow the occurrence of self-entries.
These self-entries are not created during node initialisation, but
generated based on information contained in received RREP messages.

The processing of RERR messages in AODV-UIUC and AODV-UCSB
does not adhere to the RFC specification (or even the draft versions that these implementation
are based upon).
Due to this non-adherence, we are unable to re-create the routing loop example in \SSect{self_to_loop}.
However, we note that if both AODV-UIUC and AODV-UCSB were to strictly\rule{0mm}{6.4mm} follow the RFC specification with respect to the RERR processing, loops would have been created.

Even though the routing loop example of \SSect{self_to_loop} could not be recreated,
  both implementations allow a decrease of destination sequence numbers in \rtes to occur,
  as a result of following Interpretation \ref{int2b}. 
  This contradicts the idea
  of monotonically increasing destination sequence numbers, and can give rise to routing loops in a
  straightforward way \cite[Sect.~8.1]{TR11}.

In AODV-ns2, self-entries are allowed to
occur in nodes and the processing of RERR messages
follows the RFC specification. It follows Interpretations \ref{int2a}, \ref{int3a} and \ref{int4b}.
However in AODV-ns2, whenever a node generates a RREQ message,
sequence numbers are incremented by two instead of by one as specified in the RFC\@.  We have modified the
code such that sequence numbers are incremented by one whenever a node
generates a RREQ message, and are able to replicate the routing loop example
of \SSect{self_to_loop} in the ns2 simulator, with the results
showing the existence of a routing loop between node \ns and node \nx
in Figure~\ref{fig:routingloop}(h). However, we note that even if the code
remains unchanged and sequence numbers are incremented by two, AODV-ns2 can still
yield loops; the example is very similar to the one presented and only
varies in subtle details.

In sum, we discovered not only that three out of five AODV implementations can produce routing loops, 
but also that there are essential differences in various aspects of protocol behaviour. 
This is due to different interpretations of the RFC by the developers of the AODV implementations.

\section{Methodology}\label{sec:methodology}

In Section~\ref{sec:RFCrevisited} we discussed several interpretations of the RFC\@. 
We also stated which of these interpretations are loop free and which are not.
To show that there is a possible routing loop, a single example, like the one we have given in 
Section~\ref{sec:loops}, is sufficient. 

The statement that a particular interpretation is loop free is much harder to show. 
Looking at the specification of AODV and AODV-based protocols,
loop freedom is claimed in the preamble; 
the justification is then given by a short informal statement. For example, AODV
``{\sf uses
   destination sequence numbers to ensure loop freedom at all times
   (even in the face of anomalous delivery of routing control messages),
   avoiding problems (such as ``counting to infinity'') associated with
   classical distance vector protocols.\/}''~\cite[Sect.~1]{rfc3561}.
As we have shown, these statements 
are not sufficient and not necessarily true.
The only way to guarantee loop freedom for some interpretations and to analyse 
all (reasonable) interpretations in a systematic manner is by the use of formal modelling and analysis.

\subsection{Formal Modelling and Analysis}
Ideally, any specification is free of ambiguities and contradictions. Using English prose only, this is nearly impossible to achieve. 
Hence every specification should be equipped with a formal specification. 
The choice of an appropriate specification language is often secondary, although it has high impact on the analysis. 
The use of {\em any} formal language helps to avoid ambiguities and to precisely describe the intended behaviour. 
Examples for modelling languages are\linebreak[3]
(i) the Alloy language, which is used to model Chord \cite{Zave12};\linebreak[2]
(ii) timed automata, which are the input language for the {\uppaal}
model checker, used by Chiyangwa, Kwiatkowska \cite{CK05} and others~\cite{TACAS12} to reason about AODV;
(iii) routing algebra as introduced by Griffin and Sobrinho~\cite{GS05}
or (iv) AWN, a process algebra particularly tailored for (wireless mesh) routing protocols~\cite{ESOP12,MSWIM12}.

The analysis yielding the results presented in this paper is based on a formal model using AWN\@.
The reason why we chose this formal language is two-fold: 
on the one hand it is tailored for wireless protocols and therefore offers primitives such as {\bf broadcast}; 
on the other hand, it defines the protocol in a pseudo-code  that is easily readable by
any network or software researcher/engineer.
(The language itself is implementation independent). 
Table~\ref{tb:awn} presents the main primitives of AWN;
the full specification of AODV in AWN can be found in \cite{TR11}.

\vspace{-1.7ex}
\begin{table}[ht]
\caption{ AWN language (main primitives)}
\vspace*{1.5ex}
\centering
{\footnotesize
  \setlength{\tabcolsep}{2.6pt}
 \begin{tabular}{|l|p{0.53\columnwidth}|}
\hline
\rule[6.5pt]{0pt}{1pt}%
$X(\dexp{exp}_1,\ldots,\dexp{exp}_n)$& process name with arguments\\
$P+Q$ &  choice \\
$\cond{\varphi}P$&if statement: \newline
		execute $P$ if condition $\varphi$ holds\\
$\assignment{\keyw{var}:=\dexp{exp}}P$&assignment followed by $P$\\
$\broadcastP{\dexp{ms}}.P $&broadcast \dexp{ms} followed by $P$\\
$\groupcastP{\dexp{dests}}{\dexp{ms}}.P$&iterative unicast to all destinations
	\dexp{dests}\\
$\unicast{\dexp{dest}}{\dexp{ms}}.P \prio Q$& unicast $\dexp{ms}$ to $\dexp{dest}$:\newline if successful proceed with $P$; otherwise with $Q$\\
$\deliver{\dexp{data}}.P$&deliver data to application layer\\
$\receive{\dexp{msg}}.P$&receive a message\\
$P\|Q$		&parallel composition of nodes\\
\hline
\end{tabular}}
\label{tb:awn}
\end{table}

Based on a formal specification one can now perform a careful analysis of the model to see if the
model is consistent and if it satisfies
properties such as loop freedom. Our analysis uses ``classical'' paper-and-pen verification
techniques, and, in this end, 
guarantees the loop freedom of some interpretations of the AODV specification. 
\pagebreak[4]

\subsection{Using Formal Methods to Augment RFCs}
Though we have not shown our proofs in the present paper---the purpose was to show 
that sequence numbers do not a priori guarantee loop freedom and that formal methods are needed---our analysis is based 
on a rigorous, formal and mathematical approach. 
As mentioned before, each interpretation of the RFC has been formalised in an unambiguous way.

We strongly believe that a ``good'' specification should consist of both a formal specification such as the one given in 
\cite{MSWIM12,TR11} \emph{and\/} an English description. 
The English text then serves the purpose of informing the reader about the main behaviour of the protocol and explains design decisions; 
the formal specification gives all the details, without allowing any ambiguities.

The IETF argues for the value of formal methods for specifying, analysing and verifying 
protocols. 
\begin{quote}
``{\sf Formal languages are useful tools for specifying parts of protocols. 
However, as of today, there exists no well-known language that is 
able to capture the full syntax and semantics of reasonably rich IETF protocols.\/}''
\mbox{}\hfill [IETF Web page\raisebox{3.8pt}{\footnotescriptsize{\ref{foot}}}]
\end{quote}

\noindent 
The quote is dated 1 Oct 2001; we 
believe that since then formal methods have advanced to such a state that they are now able to
capture the full syntax and the full semantics of protocols and should be used for protocol specification and analysis. 

\section{Related Work}\label{sec:related}
Analysing and verifying routing protocols has a long tradition.
Merlin and Segall \cite{MS79} were amongst the first to use sequence numbers to guarantee loop freedom of a routing protocol.
As we have pointed out, at least two proofs for AODV's loop freedom
have been proposed~\cite{AODV99,ZYZW09}. 
These proof attempts are discussed in Section~\ref{ssec:sqns}.
Besides,
other researchers
have used formal specification and analysis techniques to investigate the correctness of AODV; we point only at some examples.

A preliminary draft of AODV has been shown to be not loop free by Bhargavan et al.\
in~\cite{BOG02}. Since then, AODV has changed to such a degree that
the analysis of \cite{BOG02} has no bearing on the current version \cite{rfc3561}.
Furthermore, their loop had to do with timing issues (as is also the case in \cite{Garcia04, Rangarajan05}), whereas ours is time-independent.
In the same paper they show the use  of model checking on a draft of AODV, demonstrating the feasibility and value of automated verification of routing protocols. Using the model checker SPIN they were able to find/verify the routing loop in the preliminary draft. Musuvathi et al.~\cite{MPCED02} introduced the CMC model checker primarily to search for coding errors in implementations and used AODV as an example. Chiyangwa and Kwiatkowska~\cite{CK05} use the timing features of the model checker {\uppaal} to study the relationship between the timing parameters and the performance of route discovery. 
A ``time-free'' model of AODV is exhaustively analysed by Fehnker et
al.\ in~\cite{TACAS12}, where variants of AODV, which
yield performance improvements, are  proposed and analysed. The presented model is based on
the process algebra AWN introduced by the same authors in~\cite{ESOP12}; the complete model of AODV can be found in~\cite{TR11}.

Next to the analysis of AODV with various formal methods, 
formal languages are also used for the specification and verification of other reasonably rich protocols.
In \cite{GS05}, Griffin and Sobrinho use path algebras and  algebraic routing to 
define a metarouting language that allows the definition of routing protocols. Their main interest lies in the combination 
of different algebras with applications in interdomain routing protocols such as BGP\@. Zave et al.\ provide abstractions for implementing the 
SIP protocol; the major elements of the language are presented in~\cite{ZaveEtAl09}.

\section{Discussion \& Conclusion}\label{sec:conclusion}
We have shown that, in contrast to common belief, sequence numbers do not 
guarantee loop freedom, even if they are increased monotonically over time and 
incremented\linebreak whenever a new route request is generated. This was mainly achieved by 
analysing the Ad hoc On-demand Distance Vector (AODV) routing protocol. 
We have shown that AODV can yield routing loops.

Of course, one could argue that the given loop example does not occur in practice very often, since the 
sequence in which different requests have to be initiated and afterwards handled by different nodes
might be really rare. However, it has been claimed that routing loops are avoided in {\em all possible scenarios} by the use of 
sequence numbers---this has been proven to be incorrect. 
Here, it does {\em not} matter how often this scenario occurs in real life. The fact that the example exists and can occur, makes the protocol flawed and disproves the  fact that monotonically increasing sequence numbers are sufficient to guarantee loop freedom.

Next to that we have analysed
several different interpretations of the AODV RFC\@. 
It turned out that
several interpretations can yield unwanted behaviour such as routing loops. 
We also found that implementations of 
AODV behave differently in crucial aspects of protocol behaviour, although they all follow the lines of the RFC\@. 
This is often caused by ambiguities, contradictions or unspecified behaviour in the RFC\@.
Of course a specification ``{\sf needs to be reasonably implementation independent\/}''%
\footnote{\label{foot}\url{http://www.ietf.org/iesg/statement/pseudocode-guidelines.html}}
and can leave some decisions to the software engineer; however it is our belief that any specification should be clear and 
unambiguous enough to guarantee the same behaviour when given to different developers. 
As demonstrated, this is not the case for AODV, and likely
not for many other RFCs provided by the IETF.

Our work confirms that RFCs written merely in a natural
language contain ambiguities and contradictions. As a consequence, the
various implementations
depart in various ways from the RFC\@. Moreover, semi-informal
reasoning is inadequate to ensure critical safety-properties like
loop freedom.  We believe that formal specification languages and
analysis techniques---offering rigorous verification and ana\-lysis techniques---are
now able to capture the full syntax and semantics of reasonably rich
IETF protocols. These are an indispensable augmentation to natural
language, both for specifying protocols such as AODV, AODVv2 and HWMP,
and for verifying their essential properties.

\subsubsection*{Acknowledgement.}
NICTA is funded by the Australian Government as represented by the Department of Broadband,
Communications and the Digital Economy and the Australian Research Council through the ICT Centre of
Excellence program.

\label{lastpage}
\end{document}